\documentclass[times, review,10pt]{elsarticle}

\usepackage{amssymb}
\usepackage{graphicx}
\usepackage{color}
\usepackage{microtype}
\usepackage[english]{babel}
\usepackage{titlesec}
\usepackage{amsmath, amsthm, amsfonts, amssymb, bbold}
\usepackage{booktabs}
\usepackage{float,comment}

\usepackage{amsmath} % Pacote padrão para símbolos matemáticos avançados
\usepackage{bm}      % Pacote para negrito em modo matemático
\usepackage{braket} % Pacote para notação bra-ket
\journal{Pattern Recognition}

\begin{document}

\begin{frontmatter}

%% Title, authors and addresses

%% use the tnoteref command within \title for footnotes;
%% use the tnotetext command for theassociated footnote;
%% use the fnref command within \author or \affiliation for footnotes;
%% use the fntext command for theassociated footnote;
%% use the corref command within \author for corresponding author footnotes;
%% use the cortext command for theassociated footnote;
%% use the ead command for the email address,
%% and the form \ead[url] for the home page:
%% \title{Title\tnoteref{label1}}
%% \tnotetext[label1]{}
%% \author{Name\corref{cor1}\fnref{label2}}
%% \ead{email address}
%% \ead[url]{home page}
%% \fntext[label2]{}
%% \cortext[cor1]{}
%% \affiliation{organization={},
%%             addressline={},
%%             city={},
%%             postcode={},
%%             state={},
%%             country={}}
%% \fntext[label3]{}

\title{PUBO Formulation for MST and Application to Optimum-Path Forest}

%% use optional labels to link authors explicitly to addresses:
%% \author[label1,label2]{}
%% \affiliation[label1]{organization={},
%%             addressline={},
%%             city={},
%%             postcode={},
%%             state={},
%%             country={}}
%%
%% \affiliation[label2]{organization={},
%%             addressline={},
%%             city={},
%%             postcode={},
%%             state={},
%%             country={}}

\author[label2]{Guilherme E. L. Pexe}
\author[label3]{Lucas A. M. Rattighieri}
\author[label1]{Leandro A. Passos}
\author[label1]{Danilo S. Jodas}
\author[label1]{Douglas Rodrigues}
\author[label1]{Felipe F. Fanchini}
\author[label1]{Jo\~ao P. Papa\corref{cor1}}
\author[label1]{Kelton A. P. Costa}

\affiliation[label2]{organization={Instituto de Física de São Carlos, Universidade de São Paulo},
                     addressline={Av. Trab. São Carlense, 400},
                     city={São Carlos},
                     postcode={13560-970},
                     state={São Paulo},
                     country={Brazil}}

\affiliation[label3]{organization={Instituto de Física Gleb Wataghin, Universidade Estadual de Campinas},
                     addressline={R. Sérgio Buarque de Holanda, 777},
                     postcode={13083-859},
                     city={Campinas},
                     state={São Paulo},
                     country={Brazil}}

\affiliation[label1]{organization={São Paulo State University, School of Sciences},
                     addressline={Av. Eng. Luís Edmundo Carrijo Coube, 2085},
                     city={Bauru},
                     postcode={17033-360},
                     state={São Paulo},
                     country={Brazil}}

\cortext[cor1]{Corresponding author. Email: joao.papa@unesp.br}

%\affiliation[label2]{organization={Institute of Computational Intelligence, Czestochowa University of Technology},
 %                    addressline={Al. Armii Krajowej 36},
  %                   city={Czestochowa},
   %                  postcode={42-200},
    %                 state={Silesian Voivodeship},
     %                country={Poland}}

% First names are abbreviated in the running head.
% If there are more than two authors, 'et al.' is used.

%% Author affiliation
% \affiliation{organization={},%Department and Organization
%             addressline={}, 
%             city={},
%             postcode={}, 
%             state={},
%             country={}}

%% Abstract
\begin{abstract}
The Optimum-Path Forest is a graph-based framework for designing classifiers that exploit inter-sample connectivity. A particular variant constructs decision boundaries based on prototypes computed by a Minimum Spanning Tree (MST) over the training data, which might become prohibitive for large-scale datasets. In this context, Quantum Machine Learning has emerged as a promising approach to overcome the high computational burden of combinatorial problems. We propose a quantum-inspired approach for prototype selection in OPF classifiers by reformulating the MST problem as a Polynomial Unconstrained Binary Optimization (PUBO) task and further employing the Feedback-Based Quantum Optimization (FALQON) algorithm for Hamiltonian minimization. The PUBO formulation reduces the need for qubits and eliminates the need for auxiliary variables, thereby addressing scalability constraints in current quantum hardware. Experiments on real-world datasets demonstrate that the FALQON-optimized MST achieves accuracies comparable to those of the classical Prim's algorithm while maintaining prototype quality. While FALQON occasionally reached local minima, it did not significantly impact the accuracy of the prototype selection process.
\end{abstract}

%% Keywords

\end{frontmatter}

%% Add \usepackage{lineno} before \begin{document} and uncomment 
%% following line to enable line numbers
%% \linenumbers

%% main text
%%

%
%
\section{Introduction}
\label{s.introduction}

The emergence of quantum computing has led to significant advances in both hardware and software, particularly in quantum chips that achieve lower error rates. The convergence of quantum computing and machine learning has given rise to a new field, Quantum Machine Learning (QML), which employs quantum algorithms to enhance data analysis and pattern recognition~\cite{Aggarwal2024qml}.

Improvements in quantum hardware have enabled the development of intricate QML models, including Quantum Neural Networks~\cite{abbas2021power} and Quantum Generative Adversarial Networks~\cite{nokhwal2024quantum}, with potential advantages over their classical counterparts. Additionally, theoretical studies have shown that QML algorithms, such as the Quantum Approximate Optimization Algorithm (QAOA) and Variational Quantum Eigensolvers (VQE), can achieve exponential speedups in some scenarios, including high-dimensional data analysis and combinatorial optimization~\cite {dutta2024quantum,ye2023towards}. 

An underexplored area in QML concerns graph-based approaches. Graphs play a significant role in machine learning, notably when the dataset natively encodes entities and their relationships, such as proteins and molecules, energy distribution networks, and routing maps. The Optimum Path Forest (OPF) is a graph-based framework that models classification problems as a reward-competition task, comprising supervised~\cite{PapaIJIST:09,PapaPR:12,PapaPRL:17}, unsupervised~\cite{RochaIJIST:09}, and semi-supervised learning algorithms~\cite{amorim2014semi}. OPF has attained promising results in many applications, such as medicine~\cite{ribeiro2015unsupervised}, ensemble-based learning~\cite{jodas2023opfsemble,lucas2024ensemble}, and data imbalance~\cite{passos2022handling,jodas2022multiclass,passosSCBMS:2020}, to name a few. Its primary mechanism relies on a competition process among the so-called \emph{prototypes}, key nodes that best represent a group of samples in an embedding.

The most widely used OPF variant, likely due to its simplicity and effectiveness, is the one proposed by Papa et al.~\cite{PapaIJIST:09,PapaPR:12}. It employs a complete graph in a supervised manner, with prototypes located near the decision boundaries. They can be computed using the well-known Minimum Spanning Tree (MST), which has quadratic complexity with respect to the number of nodes (i.e., training samples). The adjacent nodes with different labels are marked as prototypes. The rationale is to place such key samples in regions more prone to misclassification, thereby \emph{preventing} samples from one class from being dominated by samples from another label during learning\footnote{All\'ene et al.~\cite{AlleneIVC:10} indicated that OPF has no classification error during learning using MST for prototype computation, provided a unique MST is present. Such a scenario occurs when all arc-weights among training samples are different.}.

Although MST computation is known to be polynomial, it is impractical for modern datasets that may contain millions of samples. Classical MST algorithms are inherently iterative and greedy, making them challenging to map directly onto quantum operations. Current quantum algorithms, such as Grover's search or quantum optimization frameworks, may not directly provide significant speedups for MST problems, as these algorithms are efficient for unstructured search or specific optimization tasks, whereas MST computation is inherently structured.

More efficient optimization methods, such as the Polynomial Unconstrained Binary Optimization (PUBO) problem, may yield better solutions for graph-related problems, including Minimum Spanning Tree algorithms. PUBO is a general optimization framework that finds optimal values for binary variables to minimize a polynomial objective function and can be applied to many problems, including combinatorial optimization tasks. Additionally, PUBO allows the inclusion of higher-order terms in the objective function, thereby overcoming the quadratic limitation of the Quadratic Unconstrained Binary Optimization (QUBO) method~\cite{SteinGECCO:23}.

In quantum computing, optimization problems are often represented by a Hamiltonian whose ground state encodes the optimal solution. By mapping the MST problem onto a suitable PUBO Hamiltonian, one can enforce constraints to avoid cycles while minimizing the total edge cost. Specifically, each edge and vertex constraint can be incorporated into polynomial terms of the Hamiltonian, ensuring that the global minimum (ground state) corresponds to a valid MST. This formulation leverages the expressive power of PUBO, which accommodates higher-order interactions required to enforce MST constraints. 

The minimization can be carried out using Feedback-based Quantum Optimization FALQON \cite{PhysRevLett.129.250502}, which was initially proposed for optimization problems and formulated using Lyapunov control principles \cite{PhysRevA.106.062414}. This approach was later extended and analyzed more generally for ground state preparation under the name Feedback-based Quantum Algorithms FQA \cite{PhysRevResearch.6.033336}. After these developments, feedback-based strategies were applied both to optimization tasks and to ground state preparation of more structured Hamiltonians, including analyses of the critical properties of the ANNNI model \cite{PhysRevB.110.224422}. From there, several variations were introduced to improve performance, including time rescaling methods designed to accelerate convergence \cite{qc91-5mj2} and a second order quadratic approximation that reduces circuit depth while preserving the feedback structure \cite{PhysRevResearch.7.013035}.

The only attempt to provide an OPF implementation in quantum computers was recently presented by Miranda et al.~\cite{OPF_tsp}, which replaced the Minimum Spanning Tree with the Traveling Salesman Problem algorithm to find prototypes. Although the proposed approach achieved similar results to classical OPF, their approach faces two primary shortcomings: (i) TSP does not guarantee a non-cycled graph, which may violate the OPF correctness, and (ii) their approach was validated with small graphs.

This work extends the approach proposed by Miranda et al.~\cite{OPF_tsp} by using PUBO with FALQON to compute the MST and the prototypes. Our approach can simultaneously optimize multiple objectives, such as minimizing classification error and maximizing the diversity of selected prototypes, thereby significantly extending OPF to quantum computing tasks. We can summarize our main contributions as follows:

\begin{itemize} 
    \item to propose a quantum alternative to the MST computation using FALQON and PUBO approaches;
    \item to propose a hybrid variant of the OPF algorithm that employs quantum computing to locate prototypes;
    \item to improve our prior work~\cite{OPF_tsp} with a more robust approach to prototype selection.
\end{itemize} 

The remainder of the paper is organized as follows. Section~\ref{s.theoretical} provides the theoretical background on the Optimum-Path Forest classifier, and Section \ref{s.proposed} details the proposed method. Section~\ref{s.methodology} describes the methodology, datasets, and experimental setup. Section~\ref{s.results} discusses the results, and the conclusions and future works are stated in Section~\ref{s.conclusions}.

\section{Theoretical Background}
\label{s.theoretical}

This section provides a theoretical background concerning the Minimum Spanning Tree problem and the Optimum-Path Forest classifier.

\subsection{Minimum Spanning Tree}
\label{ss.mst}

Let $G_w=({\cal V},{\cal E})$ be a connected graph such that ${\cal V}$ and ${\cal E}$ denote the node and edge sets, respectively, and $w:{\cal V}\times{\cal V}\rightarrow\Re^+$ is a function that weights an edge $e=\langle v_i,v_j\rangle$, such that $v_i,v_j\in{\cal V}$ and $i\neq j$. Besides, let $T_w=({\cal V^\prime},{\cal E^\prime})$ be a cycle-free graph that satisfies the following properties:

\begin{enumerate}
    \item ${\cal V}^\prime = {\cal V}$.
    \item ${\cal E}^\prime \subseteq {\cal E}$.
\end{enumerate}
In this case, $T_w$ is said to be a \emph{spanning tree} from $G$, i.e., a subgraph that contains all nodes present in $G$.

The subgraph $T_w$ naturally evolves to a \emph{minimum spanning tree} provided the following restriction is applied:

\begin{equation}
{\cal E}^\prime \in \underset{{\cal E}^\ast \subseteq {\cal E}}{\arg\min} \left\{ \sum_{e \in {\cal E}^\ast} w(e) \right\}.
\end{equation}
Essentially, the minimum spanning tree is a cycle-free subgraph that spans all nodes of the original graph, and the summation of the arc weights is minimized.

Algorithms for computing the MST play a key role in numerous practical applications, including communication networks, data clustering, network analysis, and transportation systems~\cite{Saravanan}. Follow, below, the two most widely known algorithms for computing MSTs:
\begin{itemize}
\item \textbf{Kruskal's Algorithm~\cite{kruskal-original}:} a greedy algorithm that sorts all edges in non-decreasing order of weight and adds them to the spanning tree one by one, as long as they do not form a cycle.
\item \textbf{Prim's Algorithm~\cite{prim-original}:} a greedy algorithm that starts from an arbitrary node and iteratively adds the minimum-weight edge that connects a vertex in the tree to a vertex outside the tree.
\end{itemize}
Although these algorithms have polynomial-time solutions, they are not feasible for graphs with a few million samples.

\subsection{Optimum-Path Forest}
\label{ss.opf}

The Optimum Path Forest is a graph-based machine learning framework that represents data samples as nodes in a graph, with edges that define relationships between pairs of instances. The supervised version, introduced by Papa et al.~\cite{Papa2009}, arranges the training samples into a fully connected graph, with edge weights corresponding to the distances among vertices. The model calculates the prototypes, i.e., adjacent nodes that connect samples from different classes, using the MST, which is computed by adapting Prim's algorithm. The rationale for using MST for prototype computation is that we are interested in placing key samples in regions near sample boundaries, which are known to be more prone to misclassification.

As soon as the prototypes are computed, they are stored in a set ${\cal P}$, and a competition among them takes place. The rationale is to assign an optimum cost to each remaining vertex in ${\cal V}\backslash{\cal P}$, which is calculated using the $f_{max}$ function as follows:

\begin{eqnarray}
 \label{e.fmax}
f_{max}(\langle
\bm{s}\rangle) & = & \left\{ \begin{array}{ll}
  0 & \mbox{if $\bm{s}\in {\cal P}$,} \\
  +\infty & \mbox{otherwise,}
  \end{array}\right. \nonumber \\
  f_{max}(\phi_{\bm{s}} \cdot \langle \bm{s},\bm{z}\rangle) & = & \max\{f_{max}(\phi_{\bm{s}}),w(\bm{s},\bm{z})\}, 
\end{eqnarray}
where $\bm{s}$ and $\bm{z}$ are arbitrary nodes in {\cal V}, $\phi_{\bm{s}}$ represents a path starting at a prototype in ${\cal P}$ and ending at sample ${\bm{s}}$, and $\phi_{\bm{s}} \cdot \langle \bm{s},\bm{z}\rangle$ denotes the concatenation of the path $\phi_{\bm{s}}$ with the edge $\langle \bm{s},\bm{z}\rangle$. Additionally, $f_{max}(\phi_{\bm{s}})$ indicates the maximum distance among adjacent nodes along the path $\phi_{\bm{s}}$. 

The training process assigns an optimum cost $C(\bm{z})$ to each sample $\bm{z}$ in the training set ${\cal V}$ as follows:

\begin{equation}
\label{e.conquering_function}
	C(\bm{z})  =  \min_{\forall \bm{s} \in {\cal V}}\{\max\{C(\bm{s}),w(\bm{s},\bm{z})\}\}, 
\end{equation}
where $\bm{s}$ is the instance that conquers $\bm{z}$.

The classification phase computes the distance between each test sample and each training node to determine the minimum cost, as in Equation~\ref{e.conquering_function}. The test sample is then assigned the same label as the prototype along the path that yielded the optimal cost. This procedure is repeated until all samples in the test set are classified.

\section{Proposed Method}
\label{s.proposed}

This section presents the proposed procedure that formulates the MST problem as a PUBO approach. Several QUBO formulations for the MST problem were proposed in the literature, e.g., Carvalho~\cite{carvalho2022quboformulationsnphardspanning} and Lucas~\cite{lucas2014}. In particular, the formulation proposed by Lucas addresses the issue by incorporating a degree constraint into the tree, which renders the problem NP-hard. The method employs binary variables $y_{uv}$ to indicate whether the edge $(\bm{u},\textbf{v}) \in {\cal E}$ is part of the spanning tree, and incorporates penalty terms into the cost function to ensure that the problem constraints are met and that auxiliary variables are introduced. The number of required variables is given as follows:

\begin{equation}
|V|\left(\Bigl\lfloor\frac{|V|}{2}\Bigr\rfloor+1\right) + |E|\left(2 \Bigl\lfloor \frac{|V|}{2}\Bigr\rfloor + 1\right).
\end{equation}

%For each vertex $v$, variables $x_{v,i}$ are added, which indicate whether the vertex $v \in V$ is the node at level $i$ of the tree, where $i = 0, \dots, |V|/2$. 

%The variables $x_{uv,i}$ and $x_{vu,i}$ are associated with the edges $(uv)$ and serve to define the hierarchical relationship between the vertices: if vertex $u$ is closer to the root than $v$, and $v$ belongs to level $i$, then $x_{uv,i} = 1$; otherwise, if $v$ is closer to the root than $u$, we have $x_{vu,i} = 1$. The variable $z_{v,i}$, $i = 1, \dots \Delta$ is used to count the number of edges connected to each vertex $v$, ensuring that the maximum degree of each node in the spanning tree does not exceed the established limit $\Delta$.

Further, their method introduces $x_{v,i}$ variables for each node $v \in {\cal V}$, indicating that the node stands at the $i^{th}$ level of the tree, where $i = 0, \dots, |{\cal V}|/2$. Additionally, the variables $x_{uv,i}$ and $x_{vu,i}$ are associated with the edges $(\bm{u},\textbf{v})$ and serve to define the hierarchical relationship between the nodes: if a node $\textbf{u}$ is closer to the root than a node $\textbf{v}$ and $\textbf{v}$ belongs to level $i$, then $x_{uv,i} = 1$. Otherwise, if $\textbf{v}$ is closer to the root than $u$, then $x_{vu,i} = 1$. The variable $z_{v,j}$, $j = 1, \dots \Delta$ is used to count the number of edges (i.e., the degree) connected to each node $v$, ensuring that the maximum degree of each node in the spanning tree does not exceed the established limit $\Delta$. This NP-hard variant is called the Maximum Degree Constrained Minimum Spanning Tree (MDCMST).
%If the degree constraint and, consequently, the use of the variables $z_{v,i}$ are disregarded, the formulation would solve only the MST problem, requiring a smaller number of variables. The number of required variables would be given by:
%Here, we propose a PUBO formulation for the MST problem, based on Lucas et al.~\cite{lucas2014} formulation, but using only the variables $y_{uv}$ and $x_{v,i}$, disregarding the variables $x_{uv,i}$ and $x_{vu,i}$, which results in a significant reduction in the number of variables. In this formulation, the total Hamiltonian $H$ is represented by the sum of two Hamiltonians, $H_{MST} = AH_A + BH_B$, where $H_A$ is the Hamiltonian responsible for incorporating the constraints in the form of energy penalties. It is given by:

We pose the MST problem as a PUBO formulation based on Lucas~\cite{lucas2014}, which significantly reduces the number of variables by using only $y_{uv}$ and $x_{v,i}$. This formulation represents the total Hamiltonian $H_{MST}$ by the sum of two Hamiltonians, as follows:

\begin{equation}
H_{MST} = AH_A + BH_B,    
\end{equation}
where $A$ and $B$ are ad-hoc penalty parameters, and $H_A$ is the Hamiltonian responsible for incorporating the constraints in the form of energy penalties, as follows:

\begin{equation}
\begin{aligned}
H_A =&\left(1 - \sum_{v \in V} x_{v, 0}\right)^2 
+ \sum_{v \in V} \left(1 - \sum_{i=0}^{|V|/2} x_{v, i}\right)^2 \\
+& \sum_{(u,v) \in E} y_{uv} \left(1 - \sum_{i=1}^{|V|/2} \left(x_{u, i-1}x_{v,i} + x_{v, i-1}x_{u,i}\right)\right)^2 \\
+& \sum_{u \in V} \sum_{i=1}^{|V|/2} x_{u,i} \left(1 - \sum_{(u,v) \in E} y_{uv} x_{v,i-1}\right)^2.
\end{aligned}
\end{equation}

The first term requires that the tree have exactly one root node, while the second ensures that each node belongs to exactly one level of the tree. The third term ensures that each edge of the tree connects nodes at consecutive levels, preventing cycles. The last term states that each node, except the root, must be connected to a single node at a previous level by an edge, ensuring the tree is connected. The ground state of $H_A$ is zero, indicating that all constraints have been satisfied. 

Finally, the Hamiltonian $H_B$ is responsible for minimizing the cost of the edges as follows:

\begin{equation}
H_B = \sum_{uv \in E} w(u, v) y_{uv},
\end{equation}
where $w(u,v)$ is the weight of the edge $(u,v)$. 

To ensure the problem constraints minimize the total Hamiltonian, the parameter $A$ must be chosen such that the smallest penalty introduced by $H_A$ exceeds the largest possible value of $H_B$. The term $H_B$ reaches its maximum when all edges are present in the solution.

%\begin{equation}
%\max \; H_B = \sum_{(u,v) \in E} w(u,v).
%\end{equation}

On the other hand, the smallest penalty in $H_A$ occurs when there is a single minimal constraint violation, resulting in a penalty of at least $1$. To ensure that any violation of the constraints is more costly than the minimization of $H_B$, it is sufficient to impose:

\begin{equation}
A > B \sum_{(u,v) \in E} w(u,v).
\end{equation}
This condition ensures that any solution that violates the constraints has higher energy than the best feasible solution. Thus, the MST problem can be efficiently mapped to a PUBO model, requiring $|V|(\lfloor|V|/2\rfloor+1)+|E|$ variables.

\section{Methodology}
\label{s.methodology}

This section outlines the experimental design adopted to evaluate the proposed approach. We describe the datasets, the graph design process, and the implementation details of the FALQON algorithm for extracting the MST and selecting prototypes for the OPF classifier.

\subsection{Datasets}
\label{ss.datasets}

We considered four datasets in the experimental design:

\begin{itemize}
\item Heart Disease \cite{heart_disease_45}: a real-world dataset comprising $303$ samples, each characterized by $13$ features, and classified into two categories: (i) absence and (ii) presence of heart disease.
\item Ionosphere \cite{ionosphere_52}: a dataset with $351$ instances and $34$ features, classifying radar returns from the ionosphere as (i) good (indicating ionospheric structure) or (ii) bad (signals passing through the ionosphere).
\item Lung Cancer \cite{lung_cancer_62}: a dataset with 32 instances and $56$ features used to classify three pathological lung cancer types.
\item Iris \cite{iris_53}: a real dataset with 150 samples represented by $4$ features and distributed into $3$ classes.
\end{itemize}

The selected datasets exhibit distinct characteristics in terms of the number of attributes, instances, and class distributions, enabling the evaluation of the proposed approach across various scenarios. The first three datasets (Heart Disease, Ionosphere, and Lung Cancer) were also used in~\cite{OPF_tsp}, enabling a direct comparison with previous studies. We opted for those small datasets due to the limited access to quantum computers. Still, they are not yet powerful enough to handle large-scale datasets.

%Heart Disease and Ionosphere are real-world binary classification problems with applications in the medical and signal processing fields, respectively, making them suitable for testing the method in practical contexts. 

%The Lung Cancer dataset, despite its small number of instances, presents a multiclass classification challenge in a high-dimensional feature space, assessing the approach's ability to handle more complex data. The first three datasets (Heart Disease, Ionosphere, and Lung Cancer) were also used in~\cite{OPF_tsp}, enabling a direct comparison with previous studies. Finally, the Iris dataset is a classic benchmark in machine learning, widely used to validate classification methodologies.

In addition, we conducted experiments using 8-sample subsets from each dataset, divided into four training and 4 test folds, while maintaining the class proportions. The choice of subsets with eight samples per dataset was motivated by the high computational cost of simulating the quantum circuit for FALQON on larger datasets, rendering the approach infeasible. Thus, four samples were chosen for training. A balanced division of the training and test sets (4 samples per class) was adopted to ensure consistency with previous studies~\cite{OPF_tsp}, enabling a fair comparison of the results.

\subsection{Experimental Setup}
\label{ss.experiments}

%The Figure \ref{fig:pipeline opf} illustrates the flow of the training and testing process for the OPF-based classifier utilizing the FALQON algorithm. Initially, the dataset is split into training and testing sets.

%To handle the datasets and implement the OPF classifier, we used the OPFython library \cite{rosa2021simpa}. Specifically, we employed the SupervisedOPF class, which is responsible for constructing graphs from the provided data. The edge weights were calculated based on the Euclidean distance between the feature vectors of each node, ensuring consistency with the approach of \cite{OPF_tsp}.

\begin{figure*}[!htb]
    \centering
    \includegraphics[width=\textwidth]{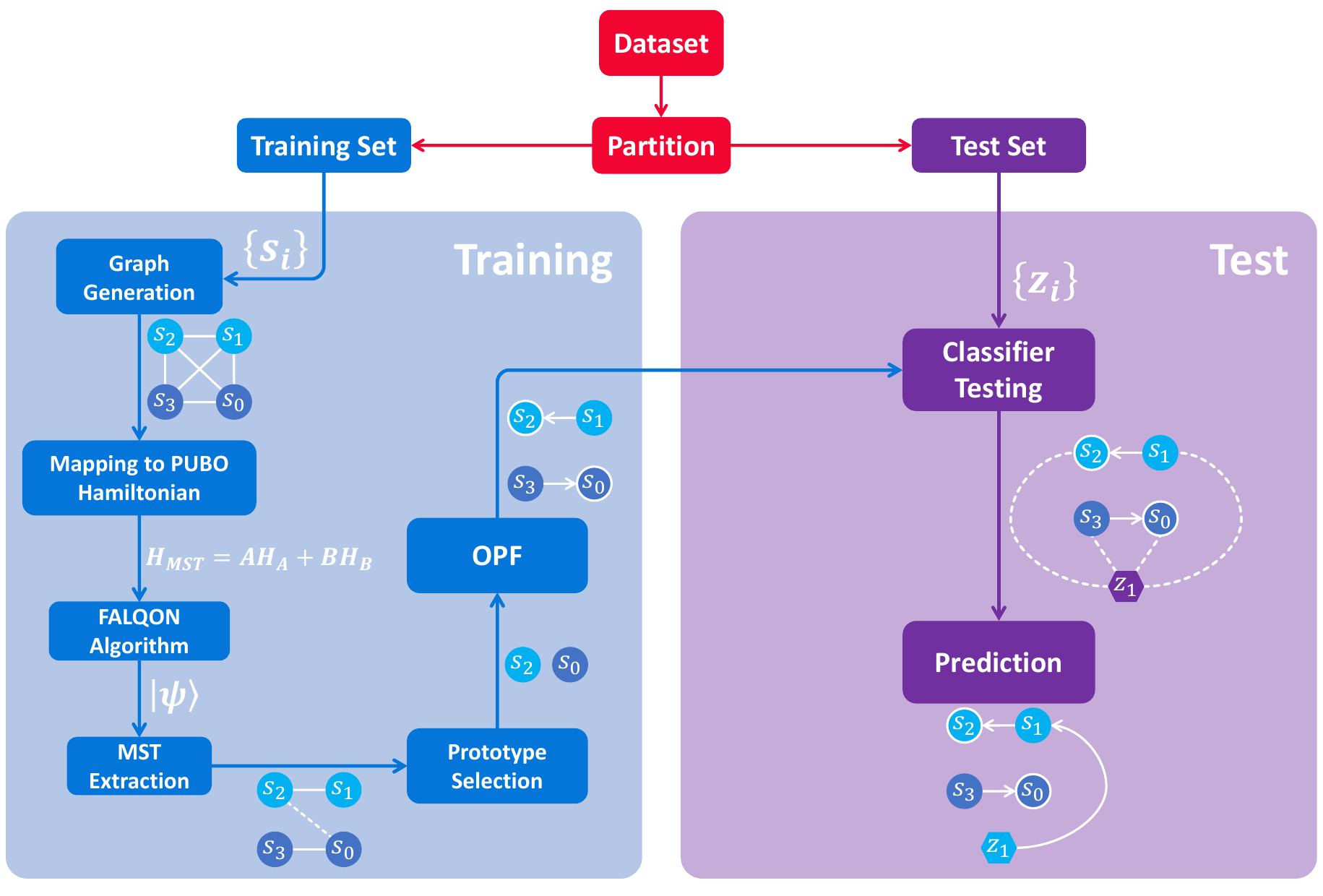}
    \caption{Flow of the training and testing process for the OPF classifier using the FALQON algorithm. The dataset is partitioned into training ($S_i$) and testing ($Z_i$) sets. During training, the data is represented as a graph and mapped to a PUBO Hamiltonian. Further, the FALQON algorithm is used to compute the MST. The selected prototypes are used to train the OPF, which is then employed to classify new examples during testing.}
    \label{fig:pipeline opf}
\end{figure*}

Figure~\ref{fig:pipeline opf} illustrates the training and testing process flow for the OPF classifier utilizing the FALQON algorithm. The process was implemented using the OPFython library~\cite{rosa2021simpa} to handle the datasets and the OPF classifier. Specifically, we employed the SupervisedOPF class to construct graphs from the provided data. The edge weights were computed as the Euclidean distance between each pair of node feature vectors, ensuring consistency with the approach proposed in~\cite{OPF_tsp}.

%The prototype selection process was restructured to use the MST, extracted from the ground state of a Hamiltonian $H_{MST}$. The PUBO Hamiltonian was implemented using the PyTorch library\footnote{https://pytorch.org/}, with the parameters $A = 0.1 \sum_{uv \in E} w(u,v)$ and $B = 0.1$. Additionally, a coefficient of 3 was applied to the first two terms of $H_A$, which ensures the presence of a root node and that each vertex belongs to a single level.

%The FALQON algorithm was used to find the ground state of the generated Hamiltonian, enabling the extraction of the MST and, consequently, the selection of prototypes. The FALQON implementation was also done in PyTorch, leveraging GPU acceleration to handle larger systems. % colocar dt e layer

The prototype selection process is restructured to use the MST, extracted from the ground state of the Hamiltonian $H_{MST}$. The PUBO Hamiltonian is implemented using the PyTorch library\footnote{https://pytorch.org/}, with $A = 0.1 \sum_{uv \in E} w(u,v)$ and $B = 0.1$. Additionally, a coefficient of $3$ is applied to the first two terms of $H_A$, which ensures the presence of a root node and that each node belongs to a single level. The FALQON algorithm is employed to find the ground state of the generated Hamiltonian, enabling the extraction of the MST and, consequently, the selection of prototypes. The FALQON implementation is also performed in PyTorch, considering GPU acceleration to handle larger systems.

%Due to the large number of sites in the formulation, we avoided the explicit storage of the Hermitian matrices associated with the Hamiltonian. Instead, we restructured the calculations to exploit the structure of these matrices. For operators represented by diagonal matrices, such as $H_p$ and $U_p$, we stored only the diagonal elements, applying them directly to the state vector. For operators without a diagonal representation, we performed the operations directly on the state vectors, using the action of these operators on the computational basis vectors. This approach allowed us to apply the time evolutions without explicitly constructing the operator matrices. This strategy significantly reduced computational requirements, enabling the simulation of larger systems without compromising efficiency.

Due to the large number of sites in the formulation, the explicit storage of the Hermitian matrices associated with the Hamiltonian was avoided. Instead, the calculations to exploit the structure of these matrices were restructured. For operators represented by diagonal matrices, such as $H_p$ and $U_p$, only the diagonal elements were stored and applied directly to the state vector. For operators without a diagonal representation, the operations were performed directly on the state vectors, using the action of these operators on the computational basis vectors. Such an approach allowed the application of time evolutions without explicitly constructing the operator matrices. This strategy significantly reduced computational requirements, enabling the simulation of larger systems without compromising efficiency.

% To avoid the explicit storage of matrices associated with the Hamiltonian, we restructured the calculations to exploit their structure. For diagonal operators, such as $H_p$ and $U_p$, we stored only the diagonal elements, applying them directly to the state vector. For non-diagonal operators, the operations are performed directly on the state vectors, using the action of these operators on the computational basis. This significantly reduced computational requirements, enabling more efficient simulations of larger systems.

%The flow of the training and testing process for the OPF-based classifier, using the FALQON algorithm, is illustrated in Figure \ref{fig:pipeline opf}. It shows the steps from dataset partitioning, graph construction, Hamiltonian mapping, MST extraction, prototype selection, and finally, classification.

%The simulations were conducted on Google Colab, a cloud-based computing environment that provides free access to hardware resources such as GPUs. The execution environment used the Ubuntu 22.04.4 LTS (Jammy Jellyfish) operating system, with Python 3.11.11 and the PyTorch 2.5.1+cu124 framework for the development and execution of the experiments.

%The hardware available on Google Colab included an Intel(R) Xeon(R) CPU, 2.20GHz processor, 12GB of RAM, and 108GB of storage. To accelerate the calculations, especially in operations involving FALQON, a Tesla T4 GPU was used, which supports CUDA version 12.4.

The simulations were conducted on Google Colab, a cloud-based computing environment that provides free access to GPU resources. The execution environment used the Ubuntu 22.04.4 LTS (Jammy Jellyfish) operating system, Python 3.11.11, and the PyTorch 2.5.1+cu124 framework to develop and execute the experiments. The hardware available on Google Colab included an Intel(R) Xeon(R) CPU, a 2.20GHz processor, 12GB of RAM, and 108GB of storage. To accelerate calculations, especially in FALQON operations, a Tesla T4 GPU with CUDA 12.4 was used.

\section{Results and Discussion}
\label{s.results}

%This section analyzes the performance of the proposed method for prototype selection in the OPF classifier. Comparisons are made between the FALQON-based approach and classical methods, evaluating the accuracy obtained on the considered datasets. Additionally, we discuss the algorithm's convergence and the impact of parameter choices on the obtained solutions.  

%All simulations were performed with a time step of $\Delta t = 0.01$ and 10,000 layers, values that proved to be consistent for the convergence of the FALQON algorithm. 

This section analyzes the performance of the proposed method for prototype selection in the OPF classifier. It compares the FALQON-based approach with classical techniques, including the one proposed by Miranda et al.~\cite{OPF_tsp}, and evaluates the accuracy obtained on the considered datasets. Additionally, it discusses the algorithm's convergence and the impact of parameter choices on the obtained solutions. All simulations were performed with a time step of $\Delta t = 0.01$ and $10,000$ layers, which were sufficient for the FALQON algorithm to converge. 

\subsection{Ground State Preparation with FALQON- PUBO}

%First, we evaluated the efficiency of FALQON in preparing the ground state of the Hamiltonian associated with the MST problem. For this, we executed the algorithm 100 times for each dataset. In each execution, we randomly selected four samples from the dataset, mapping them into a weighted graph with four vertices.

First, we evaluate the efficiency of the proposed approach for preparing the ground state (i.e., the solution) of the Hamiltonian associated with the MST problem using the proposed PUBO-based formulation. The procedure executes the algorithm $100$ times for each dataset. In each execution, it randomly selects four samples from the dataset and maps them to a weighted graph with four vertices. Table~\ref{tab:resultados_falqon} summarizes the results. 

\renewcommand{\arraystretch}{1.2} % Espaçamento entre as linhas

\begin{table}[h]
\centering
\begin{tabular}{lcc}
\toprule
Dataset      & Success Rate (\%) & Present in Top 10 (\%) \\
\midrule
Heart Disease        & 59.00 & 82.00 \\
Lung Cancer  & 46.00 & 70.00 \\
Ionosphere   & 46.00 & 58.00 \\
Iris         & 87.00 & 88.00 \\
\bottomrule
\end{tabular}
\vspace{0.4cm}
\caption{Success rate of FALQON in preparing the ground state of the MST Hamiltonian. The first column shows the percentage of executions where the most probable state corresponds to the ground state. The second column presents the frequency with which the ground state appeared among the top $10$ most probable states.}
\label{tab:resultados_falqon}
\end{table}

The performance of FALQON in preparing the ground state varies across different datasets. In the Iris dataset, the optimal solution was the most probable state in $87\%$ of executions, whereas in the Heart Disease dataset this rate was $59\%$. For the Lung Cancer and Ionosphere datasets, the ground state appeared most probable in $46\%$ of the executions. Additionally, even when the ground state was not the most probable, it was among the top $10$ measured states in up to $88\%$ of the executions (Iris), with percentages of 82\% for Heart Disease, $70\%$ for Lung Cancer, and $58\%$ for Ionosphere.

%The results indicate that FALQON frequently succeeds in preparing the ground state of the Hamiltonian, but its success rate varies across datasets. In some cases, the ground state does not appear as the most probable but is still among the most relevant, suggesting that the algorithm may be affected by local minima in the optimization landscape. This possibility is reinforced by the fact that formulating the problem as PUBO can introduce additional complexities in the Hamiltonian's structure, impacting the algorithm's convergence.

%As an example, we applied FALQON to a subset of the Heart Disease dataset. Figure  \ref{fig:grafo_mst} shows the graph constructed from this subset. In panel (a), the structure of the original graph \(G = (V, E)\) is displayed, where the vertices \(V\) represent the samples from the dataset, and the edges \(E\) are weighted according to the Euclidean distances between the vertices. In panel (b), the corresponding Minimum Spanning Tree is presented, obtained from the original graph by removing redundant edges, preserving only those that ensure connectivity with the lowest total cost.

The results indicate that FALQON frequently prepares the ground state of the Hamiltonian, but the success rate varies across datasets. In some cases, the ground state is not the most probable. However, it remains among the most relevant, suggesting that the algorithm may be trapped in local minima of the optimization landscape. This possibility is reinforced by the fact that formulating the problem as a PUBO can introduce additional complexities into the Hamiltonian structure, potentially impacting the algorithm's convergence.

\begin{figure}[!h]
    \centering
    \includegraphics[width=1\linewidth]{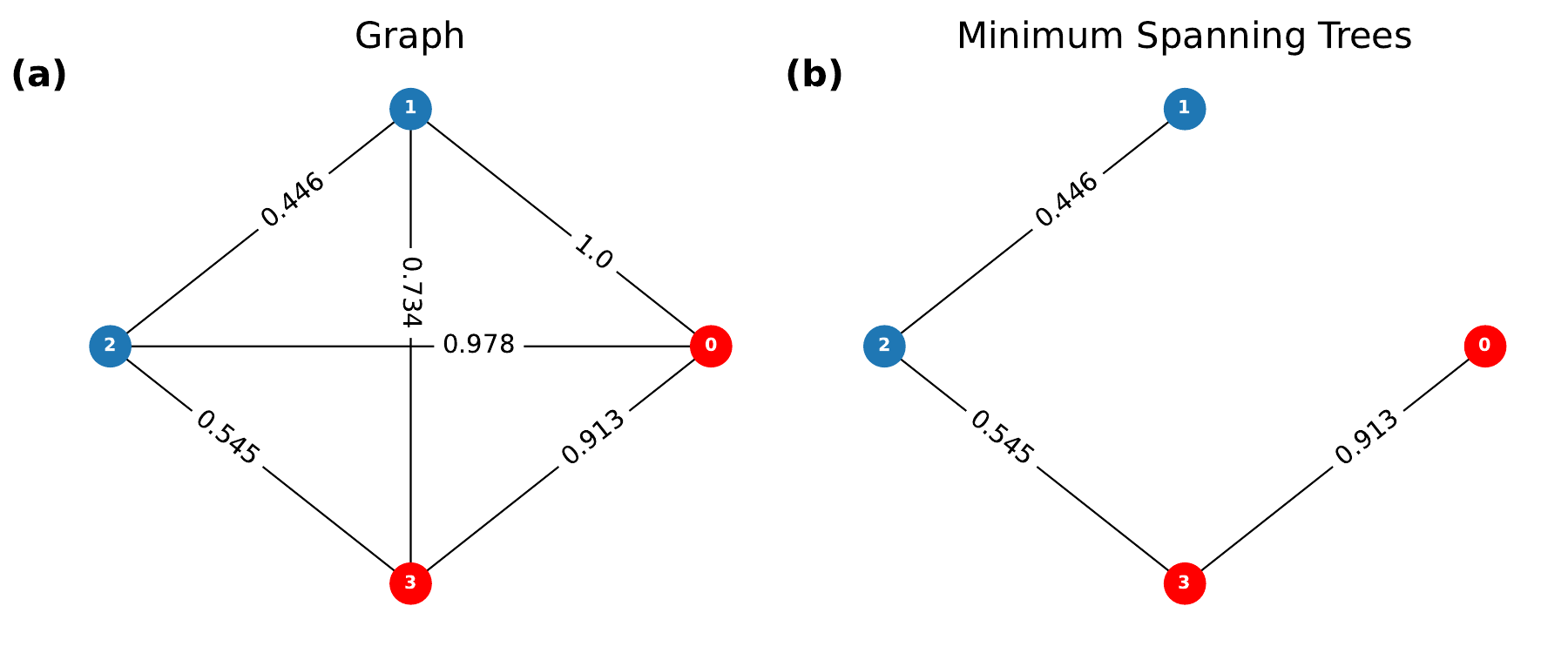}
    \caption{(a) Example graph built from the Heart Disease dataset, where the vertices correspond to the dataset samples, and the Euclidean distances between pairs of vertices weigh the edges. (b) Minimum Spanning Tree derived from the graph in (a), where the highest-cost edges are eliminated, preserving connectivity among samples with the lowest total cost.}
    \label{fig:grafo_mst}
\end{figure}

To illustrate such behavior, Figure~\ref{fig:grafo_mst} depicts the graph constructed by applying FALQON to a subset of the Heart Disease dataset. Figure~\ref{fig:grafo_mst}a displays the structure of the original graph \(G = ({\cal V}, {\cal E})\), where the edges are weighted according to the Euclidean distances between their corresponding vertices, Figure~\ref{fig:grafo_mst}b presents the corresponding Minimum Spanning Tree obtained from the original graph by removing redundant edges, preserving only those that ensure connectivity with the lowest total cost.

\begin{figure}[!h]
    \centering
    \includegraphics[width=1\linewidth]{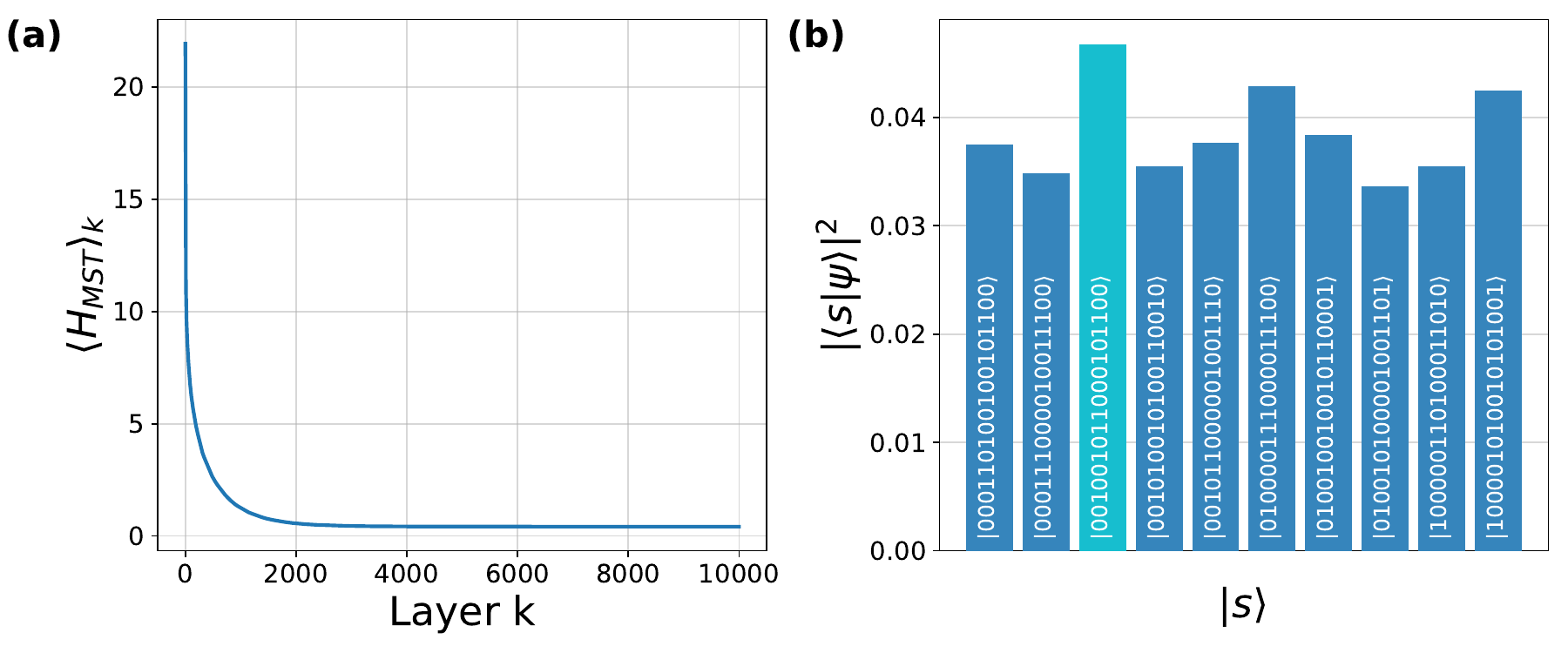}
    \caption{Results of FALQON execution on the MST problem. (a) Evolution of the expected value of the Hamiltonian $\langle H_{MST} \rangle_k$ as a function of the number of layers $k$ (\(\Delta t = 0.01\)). The monotonic decrease in energy suggests that the algorithm is approaching the ground state. (b) Probability distribution $|\langle s | \psi \rangle |^2$ of the top $10$ most probable states in the computational basis. The most probable state corresponds to the ground state of the Hamiltonian, indicating that FALQON obtained the correct solution.}
    \label{fig:la}
\end{figure}

Figure~\ref{fig:la} presents the results of FALQON execution on the graph from Figure~\ref{fig:grafo_mst}, where  Figure~\ref{fig:la}a shows the evolution of the expected value of the Hamiltonian $\langle H_{MST} \rangle_k$ as a function of the number of layers $k$, with a time step $\Delta t = 0.01$. It is observed that $\langle H_{MST} \rangle_k$ decreases monotonically as $k$ increases, indicating that the algorithm reduces the system's energy at each iteration. The initial rate of change is more pronounced but decreases as $k$ increases, suggesting possible convergence to the ground state or stagnation in a local minimum. Figure~\ref{fig:la}b shows the probability distribution $|\langle s | \psi \rangle |^2$ of the top $10$ most probable computational basis states $|s\rangle$ at the end of the execution, where FALQON obtained the state $|\psi\rangle$. The most probable state to be measured corresponds to the ground state of the Hamiltonian, indicating that the algorithm converged to the correct MST solution. However, the probability of the ground state does not reach $100\%$, suggesting the presence of competing states, possibly related to the problem's structure and the algorithm's dynamics.

\subsection{Comparing PUBO and QUBO Formulations}
\label{ss.pubo_qubo}

\begin{figure}[!h]
\centering
\includegraphics[width=1\linewidth]{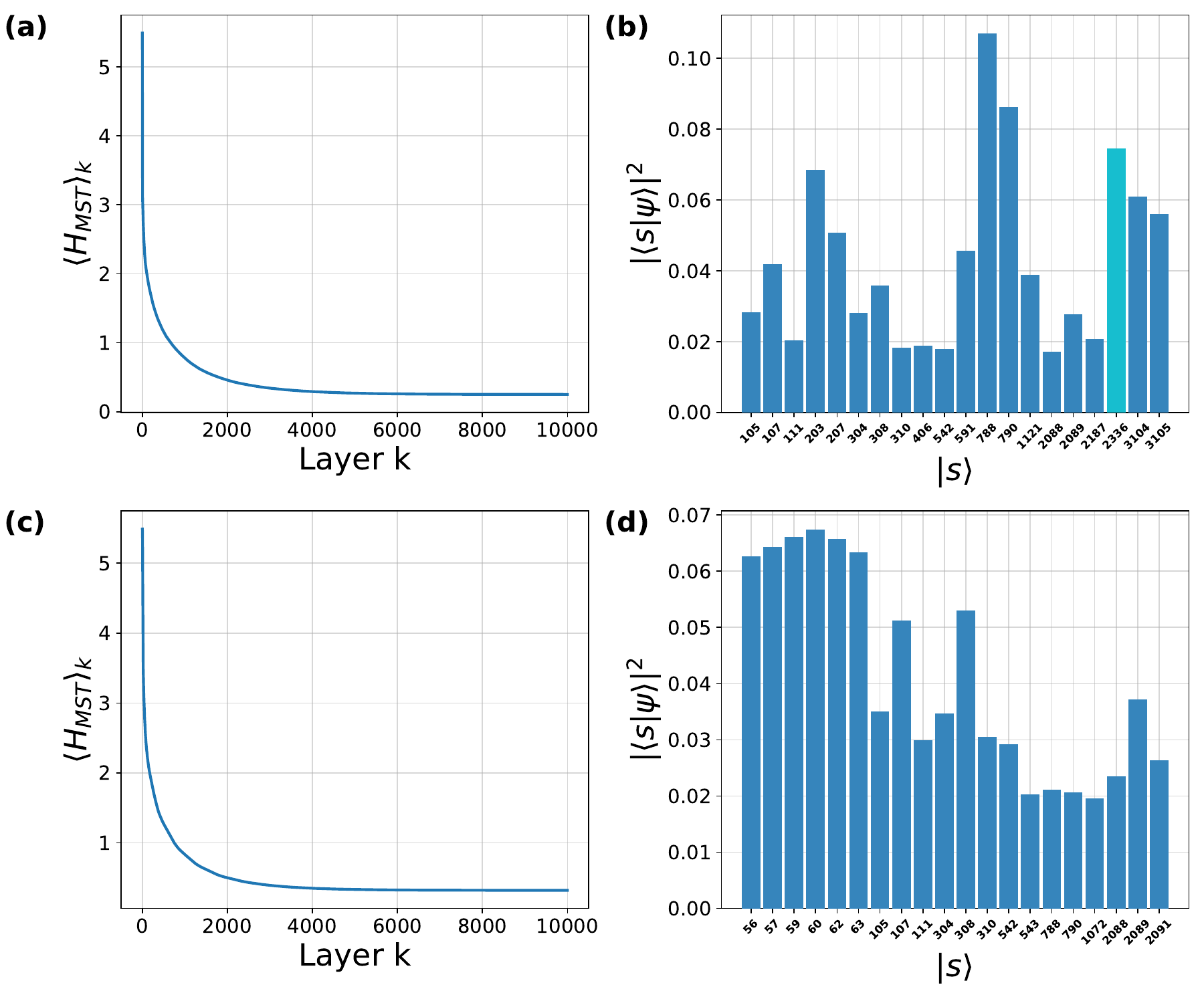}
\caption{FALQON results for the MST formulated as a QUBO problem. (a) Evolution of the expectation value $\langle H_{MST}\rangle_k$ using FALQON with different driver Hamiltonians; (b) probability distribution of the 10 most probable states for case (a), note that the solution state appears with dominant probability; (c) Evolution of the expectation value $\langle H_{MST}\rangle_k$ using standard FALQON; (d) probability distribution corresponding to case (c), in which the solution does not stand out among the measured states.}
\label{fig:fqubo}
\end{figure}

Figure~\ref{fig:fqubo} shows the results of applying FALQON to the MST task formulated as a QUBO problem, using the same parameters mentioned above (time step $\Delta t = 0.01$ and $10{,}000$ layers). Figures~\ref{fig:fqubo}a and~\ref{fig:fqubo}b correspond to the execution of FALQON when employing a strategy with different driver Hamiltonians, while Figures~\ref{fig:fqubo}c and~\ref{fig:fqubo}d present results obtained with standard FALQON (a single driver Hamiltonian).

Figures~\ref{fig:fqubo}a and c display the evolution of the expectation value $\langle H_{MST}\rangle_k$ across the layers $k$: in both cases, a monotonic decrease of the energy and convergence near the expected ground-state value are observed, indicating that the optimization protocol effectively reduces the system energy. However, the mere convergence of the expectation value \emph{does not guarantee} that the amplitude of the correct state is concentrated in a single computational-basis state. This is made explicit by comparing the probability distributions in Figures~\ref{fig:fqubo}b and~\ref{fig:fqubo}d.

Figure~\ref{fig:fqubo}b, associated with the use of multiple drivers, shows that the state encoding the MST solution emerges with significantly higher probability among the ten most likely states, i.e., in addition to lowering the energy, the dynamics generated by the varied drivers steer the amplitude toward the correct configuration. By contrast, Figure~\ref{fig:fqubo}d, which corresponds to standard FALQON, shows a final probability distribution that remains more spread across several states (without the solution state becoming dominant), even though the mean energy has also decreased.

For the QUBO formulation of the MST, the results indicate that the design of the driver Hamiltonians is critical: different drivers permit exploration of subspaces of the Hilbert space and concentration of amplitude in the desired state, whereas standard FALQON can reduce the system energy without necessarily preparing the solution state with high probability. These findings should be taken into account when comparing with the PUBO formulation, since they demonstrate that assessments based solely on energy convergence may \emph{overestimate} the method's effectiveness if the final probability distribution is not considered.

\begin{figure}[!h]
  \centering
  \includegraphics[width=\linewidth]{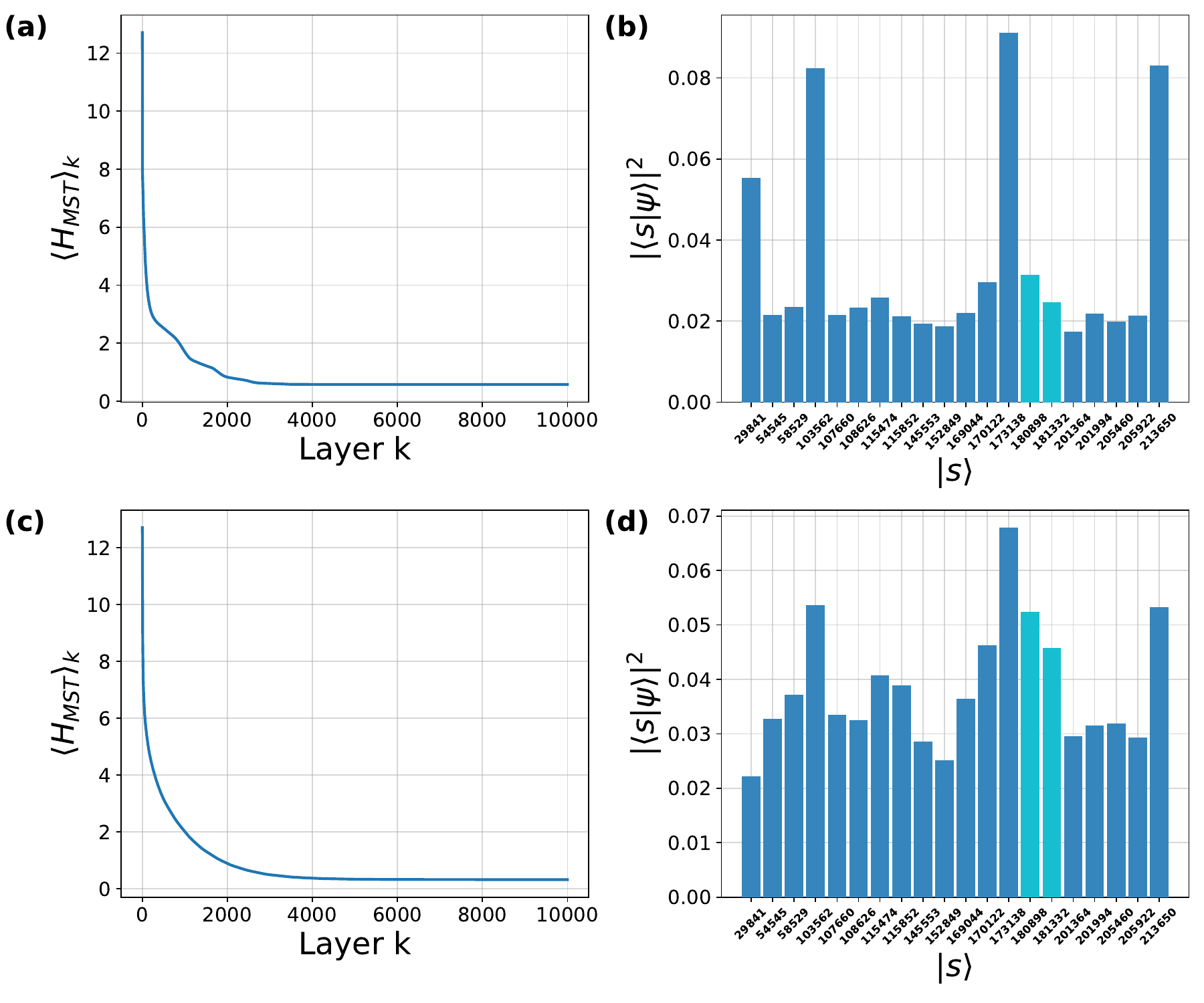}
  \caption{FALQON results for the MST formulated as a PUBO problem. (a) Evolution of the expectation value $\langle H_{MST}\rangle_k$ using FALQON with different driver Hamiltonians; (b) probability distribution of the 10 most probable states for case (a); (c) Evolution of the expectation value $\langle H_{MST}\rangle_k$ using standard FALQON; (d) probability distribution corresponding to case (c).}
  \label{fig:fpubo}
\end{figure}

Figure~\ref{fig:fpubo} shows the results for the MST formulated as a PUBO problem, obtained using the same parameters as before, and reveals a qualitatively different behavior from that observed in the QUBO case. In Figure~\ref{fig:fpubo}a and~\ref{fig:fpubo}c, one observes, as in the QUBO case, the energy curves \(\langle H_{MST}\rangle_k\) are very similar: both strategies (FALQON with multiple drivers and standard FALQON) produce a consistent decrease of the energy across the layers and converge to values close to those expected for the ground state. Therefore, at first glance, the two approaches appear equally effective in reducing energy.

The essential difference appears when comparing the final probability distributions shown in Figures~\ref{fig:fpubo}b and~\ref{fig:fpubo}c. Contrary to the behavior seen for QUBO, standard FALQON concentrates the amplitude on the solution state with a markedly higher probability among the most likely states (Figure~\ref{fig:fpubo}d). The scenario with varied drivers (Figure~\ref{fig:fpubo}b), in turn, yields a more spread-out final distribution in which the solution state does not stand out as clearly, despite having energy curves practically identical to those of the standard case. In other words, for the PUBO formulation, the energy reduction produced by different drivers does not necessarily translate into a greater concentration of amplitude on the correct configuration; the standard protocol proved more efficient at preparing the solution state with high probability.

This discrepancy highlights that the effect of the driver Hamiltonians is strongly dependent on the problem formulation. Driver Hamiltonians that help to explore useful subspaces in one formulation may, in another context, disperse amplitude into less relevant subspaces. Thus, assessments based solely on energy convergence can be misleading, especially when comparing results across different formulations, and should always be complemented by analysis of the final probability distributions. These results suggest that, for the MST in its PUBO version, the choice of drivers should be made with caution (or that the standard driver is, in this case, naturally better aligned with the topology of the solution space), and that performance evaluations should include both energy metrics and measures of the fidelity of the solution state.

\subsection{Prototype Computation Analysis}

%In this section, we compare the performance of different approaches for obtaining prototypes applied to the MST and TSP. The three methods analyzed are: (i) the traditional approach for obtaining prototypes using Prim's algorithm; (ii) the combination of FALQON with the PUBO formulation for optimizing the MST problem and obtaining prototypes; and (iii) the method proposed by Maria et al. \cite{OPF_tsp}, which uses the FALQON algorithm to optimize the Hamiltonian cycle in TSP, obtaining prototypes from this cycle.

%Table \ref{tab:accuracy} presents the average accuracies obtained for each method, considering the Heart Disease, Lung Cancer, Ionosphere, and Iris datasets. 

This section compares the performance of different approaches for obtaining prototypes using the MST and TSP: (i) the traditional approach using Prim's algorithm; (ii) the combination of FALQON with the PUBO formulation for optimizing the MST problem and obtaining prototypes; and (iii) the method proposed by Miranda et al.~\cite{OPF_tsp}, which uses the FALQON algorithm to optimize the Hamiltonian cycle in the TSP, obtaining prototypes from this cycle. 

Table~\ref{tab:accuracy} presents the average accuracies for each method across the Heart Disease, Lung Cancer, Ionosphere, and Iris datasets, with boldface results indicating the best performance. MST-Prim indicates our ground-truth results, as they are optimal and computed using classical algorithms. Performance is measured by the classification accuracy of the OPF classifier using the prototypes obtained by each method.

\renewcommand{\arraystretch}{1.2} % Aumenta levemente o espaçamento entre as linhas

\begin{table}[h!]
\centering
\resizebox{\linewidth}{!}{
\begin{tabular}{lcccc}
\toprule
Dataset & MST - Prim & MST - FALQON - PUBO (ours) & MST - FALQON -QUBO \cite{fowler2017} & TSP - FALQON~\cite{OPF_tsp} \\
\midrule
Heart Disease        & $\bm{0.70875}$ & $\bm{0.70875}$ & 0.70364 & 0.69125 \\
Lung Cancer  & $\bm{0.70390}$ & $\bm{0.70390}$ & 0.70347 & 0.70050 \\
Ionosphere   & $\bm{0.56990}$ & $\bm{0.56990}$ & $\bm{0.56990}$ & 0.55280 \\
Iris         & $\bm{0.90210}$ & $\bm{0.90210}$ & $\bm{0.90210}$ & 0.89840 \\
\bottomrule
\end{tabular}
}
\vspace{0.4cm}
\caption{OPF average accuracy obtained by each method on different datasets.}
\label{tab:accuracy}
\end{table}

Such results indicate that, in terms of accuracy, the methods using FALQON (both for MST and TSP) achieve similar performance, with a slight advantage for using Prim's algorithm to obtain the MST. However, the difference in accuracy is small, suggesting that the FALQON approach offers competitive performance compared to the traditional method. 

%On the other hand, Table \ref{tab:time} presents the average execution times for each approach. 

%When using Prim's algorithm for constructing the MST, the execution time is significantly shorter compared to the approaches using FALQON, especially for the MST case. The solution using TSP with FALQON also showed faster execution times than MST with FALQON, suggesting that the TSP method may be a more efficient choice in terms of time, albeit with a slight reduction in accuracy. It is important to note that the execution times for FALQON and TSP refer to simulations. The reason for the difference in execution times is that the TSP formulation requires fewer qubits, using six qubits, while in the MST, 18 qubits were used.

Table~\ref{tab:time} presents the average execution times for each approach. Using Prim's algorithm to construct the MST yields a significantly shorter execution time than approaches using FALQON, especially for MST construction. The solution using TSP with FALQON also achieved faster execution times than MST with FALQON, suggesting that TSP may be a more efficient choice in terms of time, albeit with a slight reduction in accuracy. It is important to note that the execution times for FALQON and TSP are reported for simulations, not on real quantum computers. The difference in execution times is that the TSP formulation requires fewer qubits than the MST. However, one should note that TSP does not guarantee an optimal solution for prototype computation with respect to the OPF classifier.

\renewcommand{\arraystretch}{1.2} % Aumenta o espaçamento entre linhas

\begin{table}[h!]
\centering
\resizebox{\linewidth}{!}{
\begin{tabular}{lcccc}
\toprule
Dataset      & MST - Prim & MST - FALQON - PUBO (ours) & MST - FALQON - QUBO \cite{fowler2017} & TSP - FALQON~\cite{OPF_tsp} \\
\midrule
Heart        & 0.027 & 82.92 & 44.03 & 9.19 \\
Lung Cancer  & 0.031 & 83.18 & 41.22 & 8.57 \\
Ionosphere   & 0.028 & 83.03 & 41.57 & 9.31 \\
Iris         & 0.020 & 83.09 & 41.42 & 8.99 \\
\bottomrule
\end{tabular}
}
\vspace{0.4cm}
\caption{Average execution time (in seconds) for different prototype generation methods.}
\label{tab:time}
\end{table}

In summary, while the traditional Prim's method is more efficient in terms of time, the FALQON-based approaches are competitive. They offer good accuracy but with higher computational costs. The choice between these approaches will depend on the specific requirements of each application, such as the need for precision or computational efficiency. In summary, while the traditional Prim's method is more time-efficient, the FALQON-based approaches are competitive. They offer good accuracy but with higher computational costs. The choice between these approaches will depend on the specific requirements of each application, such as the need for precision or computational efficiency. Such outcomes highlight the promising directions that quantum computing is taking despite its limited computational capacity.

\section{Conclusions}
\label{s.conclusions}

This work proposes a quantum-inspired approach to prototype selection for the OPF classifier by casting the Minimum Spanning Tree problem as a PUBO formulation. The associated Hamiltonian was minimized using the FALQON algorithm, allowing for the extraction of the MST and the subsequent selection of prototypes. Experiments on the Heart, Lung Cancer, Ionosphere, and Iris datasets demonstrated that the quantum approach achieves accuracies comparable to those of classical methods, such as Prim's algorithm, and to those obtained using the Hamiltonian cycle (TSP) formulation, albeit with significantly higher computational time.

Despite its promising results, the approach faces scalability challenges due to current restrictions on the number of available qubits and the high computational cost of executing FALQON. Additionally, ensuring convergence to the ground state while avoiding local-minima traps requires careful tuning of the algorithm's parameters and balancing quantum exploration with computational efficiency to enable practical application to higher-dimensional problems.

As a potential direction for future research, refining the PUBO formulation to reduce the number of required variables could reduce the demand for quantum resources. Investigating alternative quantum optimization strategies may further reduce execution times and enhance the robustness of achieving the ground state. Additionally, integrating the approach with emerging quantum hardware and developing hybrid algorithms that combine quantum and classical processing could broaden its applicability to real-world data analysis and pattern recognition.

\section{Acknowledgements}

The authors are grateful to the São Paulo Research Foundation (FAPESP) grants 2024/00998-6, 2025/07171-2, 2025/13172-1, 2023/12830-0, to the Brazilian National Council for Scientific and Technological Development grant 308529/2021-9, and to the Office of Naval Research (ONR)/Air Force Office of Scientific Research (AFOSR) grant N62909-24-1-2012.

\bibliographystyle{elsarticle-num} 
\bibliography{Reference} % Substitua "reference" pelo nome real do seu arquivo .bib, sem a extensão .bib

\end{document}